\documentclass[twocolumn]{aastex631}
\usepackage{amsmath, mathrsfs}
\usepackage{multirow}

\newcommand{\talpha}{{\tilde{\alpha}}}

\newcommand{\tgamma}{{\tilde{\gamma}}}
\newcommand{\tomega}{{\tilde{\omega}}}
\newcommand{\tepsilon}{{\tilde{\epsilon}}}
\newcommand{\calA}{{\mathcal{A}}}
\newcommand{\calH}{{\mathcal{H}}}
\newcommand{\calM}{{\mathcal{M}}}
\newcommand{\calN}{{\mathcal{N}}}
\newcommand{\calR}{{\mathcal{R}}}
\newcommand{\rmcz}{_{{\rm c},z}}
\newcommand{\rmcr}{_{{\rm c,r}}}
\newcommand{\rmc}{_{\rm c}}

\newcommand{\rmr}{_{\rm r}}
\newcommand{\orb}{_{\rm orb}}
\newcommand{\obs}{_{\rm obs}}
\newcommand{\snr}{{\rm S/N}}
\newcommand{\rhoth}{{\rho_{\rm th}}}
\newcommand{\bulge}{_{\rm bulge}}
\newcommand{\gal}{_{\rm gal}}
\newcommand{\bbh}{_{\rm BBH}}
\newcommand{\bh}{_{\rm BH}}

\newcommand{\yr}{{\rm\,yr}}

\newcommand{\iyr}{{\rm\,yr^{-1}}}
\newcommand{\hz}{{\rm\,Hz}}
\newcommand{\nhz}{{\rm\,nHz}}

\begin{document}

\title{
Constraining the origin of the nanohertz gravitational-wave background by pulsar
timing array observations of both the background and individual supermassive
binary black holes
}
\shorttitle{Origin of the nanohertz gravitational-wave background}
\shortauthors{Chen, Yu, \& Lu}

\author[0000-0001-5393-9853]{Yunfeng Chen} 
\affiliation{School of Astronomy and Space Science, University of Chinese
Academy of Sciences, Beijing 100049, China}
\affiliation{National Astronomical Observatories, Chinese Academy of Sciences,
Beijing, 100101, China} 

\author[0000-0002-1745-8064]{Qingjuan Yu}
\affiliation{Kavli Institute for Astronomy and Astrophysics, and School of
Physics, Peking University, Beijing, 100871, China}
\email{yuqj@pku.edu.cn}

\author[0000-0002-1310-4664]{Youjun Lu} 
\affiliation{National Astronomical Observatories, Chinese Academy of Sciences,
Beijing, 100101, China} 
\affiliation{School of Astronomy and Space Science, University of Chinese
Academy of Sciences, Beijing 100049, China}
\email{luyj@nao.cas.cn}

\correspondingauthor{Qingjuan Yu}

\begin{abstract}
The gravitational waves (GWs) from supermassive binary black holes (BBHs) are
long sought by pulsar timing array experiments (PTAs), in the forms of both a
stochastic GW background (GWB) and individual sources. The evidence for a GWB
was reported recently by several PTAs with origins to be determined. Here we use
a BBH population synthesis model to investigate the detection probability of
individual BBHs by the Chinese PTA (CPTA) and the constraint on the GWB origin
that may be obtained by PTA observations of both GWB and individual BBHs. If the
detected GWB signal is entirely due to BBHs, a significantly positive redshift
evolution ($\propto(1+z)^{2.07}$) of the mass scaling relation between
supermassive black holes and their host galaxies is required. In this case, we
find that the detection probability of individual BBHs is $\sim85\%$ or 
64\% if using a
period of $3.4$-year CPTA observation data, with an expectation of $\sim1.9$ or
$1.0$ BBHs detectable with a signal-to-noise ratio $\geq3$ or $5$, and it is
expected to increase to $>95\%$ if extending the observation period to $5$ years
or longer. Even if the contribution from BBHs to the GWB power signal is as
small as $\sim10\%$, a positive detection of individual BBHs can still be
expected within an observation period of $\sim10$\,years. A non-detection of
individual BBHs within several years from now jointly with the detected GWB
signal can put a strong constraint on the upper limit of the BBH contribution to
the GWB signal and help identify/falsify a cosmological origin. 
\end{abstract}

\keywords{
black hole physics (159); 
gravitational waves
(678); gravitational waves astronomy (675); pulsars (1306); supermassive black
holes (1663)}

\section{Introduction}
\label{sec:Introduction}

The gravitational wave (GW) radiation from the cosmic population of supermassive
binary black holes (BBHs) forms a stochastic GW background (GWB), which is
longly expected to be detected by the pulsar timing array (PTA) experiments. The
evidence of the presence of a GWB at the nanohertz band has been recently
reported by several PTAs, including the North American Nanohertz Observatory for
Gravitational Waves \citep[NANOGrav;][]{NG23hd}, 
the collaboration of the European PTA and the Indian PTA (EPTA+InPTA;
\citealt{EPTA23hd}),
the Parkes PTA \citep[PPTA;][]{PPTA23hd}, and the Chinese PTA
\citep[CPTA;][]{CPTA23hd}, with a confidence level of $\sim 2$-$4.6\sigma$.
However, the origin of this signal is still not determined, yet. 
The detected signal can be consistent with that predicted from 
the cosmic BBH origin, considering of the uncertainties in the model estimation
(e.g., \citealt{CYL23cgws, Muhamed23, NG23constraint, EPTA23constraint,
Curylo23, Becsy23}). It could also be (partly) originated from the cosmic
strings, phase transition, domain walls, primordial black holes, or other
processes in the early universe, which is intensively discussed recently in the
literature (e.g., \citealt{BianLG23, NG23alter, EPTA23constraint, Ellis23}). 

The GWB of the astrophysical origin may be different from the GWB of the
cosmological origin at least in the following two aspects. 
First, the former one results in a characteristic strain spectrum following the
canonical power law with a slope of $-2/3$ in the nanohertz band,\footnote{
Note that at low frequencies (e.g., $f\lesssim 10^{-9}\hz$), the
GWB may deviate from the canonical $-2/3$ power law, considering
that the environmental coupling effect \citep{BBR80, Kocsis11} and the orbital
eccentricities of cosmic BBHs  \citep{Enoki07} both lead to the bending of the
GWB spectrum \citep{ChenSY17, Rasskazov17,CYL20bbh}.
} while the latter one may
result in a GWB much different from the $-2/3$ power-law, depending on the
detailed models \citep{NG23alter, Ellis23}. 
With the accumulation of the PTA observation time, the GWB spectrum may be
accurately measured in the future and thus used to distinguish these two
scenarios. Second, the former one is fluctuating significantly at high
frequencies ($\gtrsim 10^{-8}$\,Hz) due to the small number statistics of
individual BBHs \citep{Sesana08, Roebber16, CYL20bbh}, while the latter one not.
It is expected that the signals of some individual BBHs can be loud enough to
stand out from the GWB composed of numerous weaker sources (e.g.,
\citealt{Rajagopal95, Sesana09, Ravi15, Rosado15, Mingarelli17, Kelley18,
CYL20bbh, Gardiner23, Valtolina23}). If the GWB signal detected by the current
PTA experiments is indeed originated from BBHs, there would be some loud
individual BBHs hiding in the data. Detection of any such individual BBHs by PTA
experiments would provide a consistency check for the BBH origin and/or
constraint on the contribution fraction of BBHs to the GWB, though currently no
evidence for individual BBHs was found in the data sets of NANOGrav
\citep{NG23indv} and EPTA \citep{EPTA23indv}. 

In this paper, we forecast the detection of individual BBHs by PTA experiments,
especially CPTA, with consideration of the constraint from the GWB signal
reported recently on the cosmic BBH model, and investigate the possibility to
constrain the origin of the GWB jointly by the PTA detection of individual BBH
sources and the GWB. The paper is organized as follows. In
Section~\ref{sec:BBHmodel}, we first briefly introduce the population synthesis
model for cosmic BBHs and constrain the model by using the GWB spectrum detected
recently by PTA experiments, and then describe the method to generate mock
samples for cosmic BBHs. With these samples, we investigate the detection
prospects of individual BBHs for CPTA. Our main results are presented in
Section~\ref{sec:detectability}, and the main conclusions are summarized in
Section~\ref{sec:Conclusions}.

\section{BBH population model}
\label{sec:BBHmodel}

We adopt the BBH population model constructed in \citet{CYL20bbh} (hereafter
CYL20) to estimate the GWB spectrum and generate mock cosmic BBH
populations.  We briefly describe the model as follows (see also Sections~2-3
of CYL20 for details). The model is constructed on a set of
astrophysical ingredients, including the galaxy stellar mass function (GSMF),
the galaxy merger rate, the MBH--host galaxy scaling relation, the orbital
evolution of BBHs within their merged host galaxies. These ingredients can be
obtained from either observations or numerical simulations.  Through the
incorporation of these ingredients in the model, the statistical distributions
of cosmic BBHs, their coalescence rates and gravitational wave radiation
strength can be obtained.

The cosmic BBH distribution function $\Phi\bbh(M\bh,q\bh,a,z)$, which is
defined so that $\Phi\bbh(M\bh,q\bh,a,z)dM\bh dq\bh da$ is the comoving number
density of BBHs at redshift $z$ with total mass in the range $M\bh\rightarrow
M\bh+dM\bh$, mass ratio in the range $q\bh\rightarrow q\bh+dq\bh$, and
semimajor axis in the range $a\rightarrow a+da$, can be
connected with the above model ingredients through the following
equation (cf.\ Eq.~17 in CYL20):
\begin{eqnarray}
 \Phi\bbh(M\bh,q\bh,a,z)  \quad \quad \quad \quad \quad \quad \quad \quad \quad \quad \quad \quad \quad & \nonumber \\
=  \frac{1}{N} \sum_{i=1}^{N}
\int \int d M\gal d q\gal n\gal(M\gal,z_i)
\calR\gal(q\gal,z_i|M\gal) & \nonumber \\ 
 \times p\bh(M\bh,q\bh|M\gal,q\gal,z_i) 
H(t-\tau_{a,i}) & \nonumber \\
\times \left| \frac{da_{i|M\gal,q\gal,M\bh,q\bh}}{d\tau}\right|^{-1}_{\tau
=\tau_{a,i|M\gal,q\gal,M\bh,q\bh}}, & \nonumber \\
\label{eq:phibbhi}
\end{eqnarray}
where $t$ is the cosmic time at redshift $z$, $n\gal(M\gal,z)$ is the GSMF defined so that $n\gal(M\gal,z) dM\gal$
represents the comoving number density of galaxies at redshift $z$ with stellar
mass within the range $M\gal \rightarrow M\gal+dM\gal$,
$\calR\gal(q\gal,z|M\gal)$ is the merger rate per galaxy (MRPG) defined so that
$\calR\gal(q\gal,z|M\gal) dt dq\gal$ represents the averaged number of galaxy
mergers with mass ratio in the range $q\gal \rightarrow q\gal+dq\gal$ within
cosmic time $t \rightarrow t+dt$ for a descendant galaxy with mass $M\gal$, the
BBH systems ($i = 1, 2, ..., N$) with total mass $M\bh$ and mass ratio
$q\bh$ are generated by the Monte-Carlo method according to the properties of
the merged galaxies with total mass $M\gal$ and mass ratio $q\gal$, $a_i(\tau)$
represents the semimajor-axis evolution of the BBH system $i$ as a function of
the
period $\tau$ taken since the galaxy merger, $\tau_{a,i}$ is the period taken
for the BBH semimajor-axis to decay to the value of $a$, $H(t-\tau_{a,i})$ is a
step function defined by $H(t-\tau_{a,i})=1$ if $t>\tau_{a,i}$ and
$H(t-\tau_{a,i})=0$ if $t\le\tau_{a,i}$, $z_i$ is the corresponding redshift of
cosmic time $t-\tau_{a,i}$, $p\bh(M\bh,q\bh|M\gal,q\gal,z)$ denotes the
probability distribution of the total masses and mass ratios $(M\bh,q\bh)$ of
BBHs within a galaxy merger remnant characterized by $(M\gal,q\gal)$ at
redshift $z$ and can be derived from the MBH--host galaxy relations.
In Equation (\ref{eq:phibbhi}), we ignore multiple galaxy major mergers that
could occur before the BBH coalescence since their host galaxy merger, i.e.,
setting the term $P_{\rm intact}=1$ in Equation (17) of CYL20, which is
plausible as the GWB at the PTA band is mainly contributed by galaxy or BBH
mergers within redshift lower than 2 (see Fig.~21 in CYL20).

The BBH coalescence rate  $R\bh(M\bh,q\bh,z)$, which is defined so that
$R\bh(M\bh,q\bh,z) dt dM\bh dq\bh$ represents the comoving number density of
BBH coalescences occurred during the cosmic time $t \rightarrow t+dt$, with
total BH mass within the range $M\bh \rightarrow M\bh+dM\bh$ and mass ratio
within the range $q\bh \rightarrow q\bh+dq\bh$, can be obtained through
the following equation (cf.\ Eq.~22 in CYL20):
\begin{eqnarray}
& R\bh (M\bh,q\bh,z(t)) 
= \frac{1}{N} \sum_{i=1}^{N} \iint dM\gal dq\gal \nonumber \\
& \times n\gal(M\gal,z_i) \calR\gal(q\gal,z_i|M\gal)  \nonumber \\
& \times p\bh(M\bh,q\bh|M\gal,q\gal,z_i)H(t-\tau_{a=0,i}).
\label{eq:Rbh}
\end{eqnarray}

The characteristic strain amplitude of the stochastic GWB in the PTA band,
$h\rmc$, produced by a cosmic population of BBHs at the GW frequency $f$ (in the
observer's rest frame) can be estimated by
\begin{eqnarray}
h\rmc^2(f)\simeq && \frac{4}{\pi} \frac{G}{c^2}f^{-2} 
\iiint dz dM\bh dq\bh \left|\frac{dt}{dz}\right| \nonumber\\
&& \times R\bh(M\bh,q\bh,z) \frac{1}{1+z} \left|\frac{dE}{d\ln f\rmr}\right|,
\label{eq:GWB}
\end{eqnarray}
where $c$ is the speed of light, $G$ is the gravitational constant, 
$f\rmr= (1+z)f$ is the frequency of the GW signal in the source's rest frame,
$E$ is the orbital energy of the BBH,
and $|dE/d\ln f\rmr|$ is the GW energy per unit logarithmic rest-frame
frequency radiated by an inspiraling BBH with parameters $(M\bh, q\bh, f\rmr)$
(see Eqs.~30--34 in CYL20 and also the derivation of \citealt{Phinney01}). Note
that compared with Equation~(33) in CYL20,
we set that the BBH evolution is at the gravitational radiation stage in
Equation (\ref{eq:GWB}),
as we focus on the PTA band, where $f$ is greater than the turnover frequency of
the expected GWB spectrum shown in Fig.~19 in CYL20 and the coupling of the BBH orbital
evolution with surrounding environment is negligible.

The GWB strain estimated from Equation (\ref{eq:GWB}) may suffer from
uncertainties in the involved model ingredients. 
CYL20 analyzes the effect on the estimation due to the uncertainty in
each of the model ingredients (see Section~6 therein), and find that the
uncertainty of the estimated GWB amplitude is dominated by the variation of the
MBH--host galaxy scaling relation ($\sim 1$\,dex), as compared with those due
to the MRPG ($\sim 0.3$\,dex), and that an ignoration of the time delay
$\tau_{a=0,i}$ in Equation (\ref{eq:Rbh}) can lead to an increase of the GWB
strain estimation by $\sim 0.15$\,dex. The gas effect in the BBH evolution is
neglected in the estimate of the GWB strain at the PTA band.

The BBH population model may be constrained by the detected GWB signal if it is
fully contributed by cosmic BBHs as demonstrated in \citet{CYL23cgws}, where
the common uncorrelated red noise (CURN) signal obtained in \citet{NG20cps}
is adopted.
\citet{CYL23cgws} investigate the constraint on the MBH--host galaxy scaling
relation, 
formulated in a general form of
\begin{eqnarray}
\log_{10}\left(\frac{M_{\rm BH}}{M_\odot}\right) &=&
\talpha\log_{10}\left(\frac{M\bulge}{10^{11}M_\odot}\right)+\tgamma \nonumber\\
&+&\tomega\log_{10}(1+z)+\calN(0,\tepsilon),
\label{eq:scaling}
\end{eqnarray}
as its variation dominates the uncertainty of the GWB strain amplitude
estimation.
In Equation (\ref{eq:scaling}), $M\bulge$ represents the mass of the spheroidal
components of the host galaxies (i.e., elliptical galaxies themselves or bulges
in spiral galaxies, throughout this work we use ``bulge'' to represent both
cases), the term $\tomega\log_{10}(1+z)$ describes the
redshift evolution of the scaling relation, and the term $\calN(0,\tepsilon)$
represents a random value following a normal distribution with zero mean and
standard deviation $\tepsilon$ (the ``intrinsic scatter'').
The other model ingredients are fixed by adopting the GSMF from \citet{Behroozi19} and the MRPG
from \citet{RodriguezGomez15}, and converting the host galaxy mass to the bulge
mass through the prescription in \citet{Ravi15}. The dynamical evolution of
individual BBHs is mainly based on \citet{Yu02}. 
To produce a GWB with the modelled strain amplitude being the same as the CURN
signal, it requires either (i) a positive redshift evolution of the
MBH--host galaxy scaling relation (the best fit of $\tomega=1.99$ in Eq.~\ref{eq:scaling}, see
also \citealt{Muhamed23}), or (ii) 
a normalization (the best fit of $\tgamma=9.55$ if fixing $\tomega=0$ in Eq.~\ref{eq:scaling}) much larger than the
empirically determined values (e.g., $\tgamma=8.46$ in \citealt{MM13} and
$\tgamma=8.69$ in \citealt{KH13}, which are among the largest ones for the
scaling relation determined in the literature using local MBHs with
dynamical mass measurements).
Variations of the other model ingredients can alleviate the requirements to
some extent. For example, if ignoring the time delays between BBH coalescences
and their host galaxy mergers, either the constraint of the best-fit
$\tomega=1.26$ or the best-fit $\tgamma=9.13$ is obtained (see Table~3 of
\citealt{CYL23cgws}); if the MRPG is increased by a factor of $3$, the
constraint becomes $\tomega=0.93$ or $\tgamma=9.06$. As seen from the examples,
a positive redshift evolution of the MBH–host galaxy scaling relation is still
needed or the above required best-fit normalizations of $\tgamma$ are still
higher than the largest empirically determined ones.
Here we revisit the constraint on the MBH-host galaxy scaling relation by using
the latest NANOGrav 15-year free-spectrum data set, which was derived by
considering simultaneously the HD-correlated component together with the
monopole-correlated, the dipole-correlated and the CURN components \citep[][see
the HD-w/MP+DP+CURN model therein]{NG23hd, NG23constraint}. 
The newly reported HD signal has an amplitude somewhat larger than the CURN
signal found in the NANOGrav $12.5\yr$ data set, i.e., the characteristic strain
amplitude at $f=1\,{\rm yr}^{-1}$ for the HD signal is $A_{\rm yr}=2.4\times
10^{-15}$, while the amplitude for the CURN signal is $A_{\rm yr}=1.92\times
10^{-15}$ \citep{NG20cps}.
As mentioned in option (ii) above, adopting a redshift-independent scaling
relation in Equation (\ref{eq:scaling}) yields a normalization of the scaling
relation significantly larger than those empirically determined values given by
the local MBHs with dynamical mass measurements (even being around the boundary
of the $3\sigma$ deviation from the maximum empirically determined value of
\citealt{KH13}) and the newly obtained higher $A_{\rm yr}$ from
the NANOGrav 15-year data implies a higher or more significantly deviated
normalization of $\tgamma$,
thus in this study we consider option (i), i.e., the redshift
evolution of the scaling relation by fixing the normalization $\tgamma=8.69$ as measured in
\citet{KH13}.
In the model fitting, we use the leftmost $5$ frequency bins of the
NANOGrav 15-year free-spectrum data set \citep{NG23hd}, in which the
data provide strong constraints on HD-correlated posteriors \citep{NG23constraint},
assuming that the GWB posteriors detected in the different frequency bins are
independent of each other. Note that the usage of the leftmost $5$ frequency
bins is different from \citet{CYL23cgws}, in which only the strain amplitude
at a single frequency $f=1\,{\rm yr}^{-1}$ is used for the model calibration.
We find that the parameters in Equation~\eqref{eq:scaling} that best match the
GWB spectrum data are $\talpha=1.00\pm 0.12$, $\tomega=2.07\pm 0.47$ and
$\tepsilon=0.30\pm 0.17$, suggesting again that the recent PTA observations
require a positive redshift evolution of the MBH--host galaxy scaling relation
(i.e., $\tomega=2.07$) given the adopted model settings.
This result is also roughly consistent with the recent James Webb Space Telescope (JWST) observations of MBHs at
$z\sim 4-7$ active galaxies, which suggest that the mass ratios of the MBHs to their host
galaxies at high redshift is substantially larger than those at nearby universe \citep{Pacucci23} (see also other 
works, e.g., \citealt{McLure06, Merloni10, ZLY12}).

\begin{figure*} 
\centering
\includegraphics[width=0.9\textwidth]{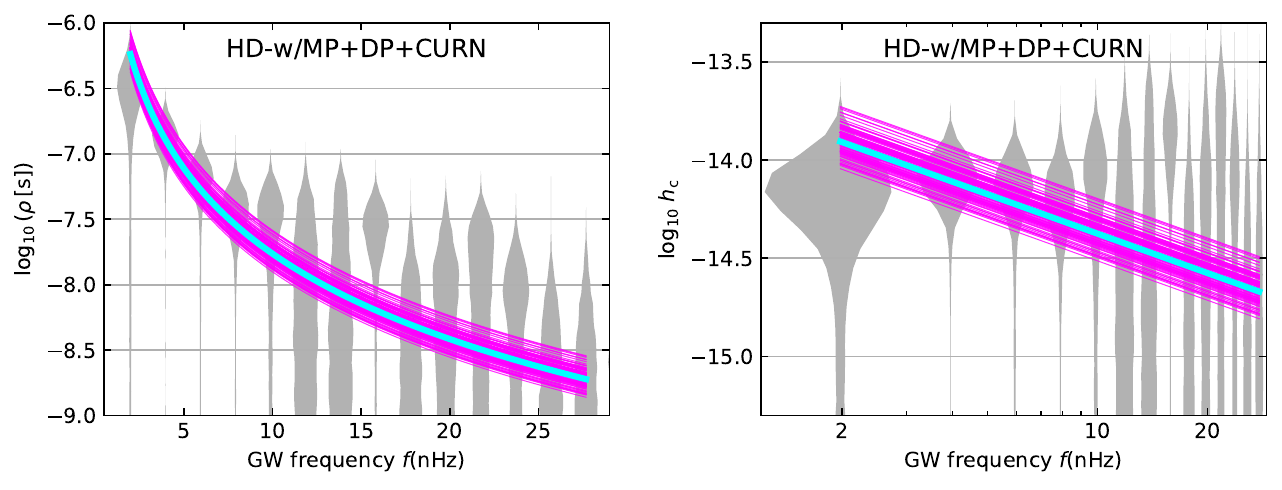}
\caption{Reconstructed GWB spectra from the BBH population model constrained by
the NANOGrav 15-year free-spectrum data (cyan and magenta lines). The grey
violins are from \citet{NG23hd} and \cite{NG23constraint}, showing the square
root of the cross-correlated timing residual power ($\rho$) in the left panel
and the characteristic strain amplitude $h\rmc$ in the right panel,
respectively, for the NANOGrav 15-year free-spectrum GWB posteriors obtained by
assuming the HD-w/MP+DP+CURN model. The cyan line in each panel shows the
reconstructed GWB spectra for the reference BBH population model, while the thin
magenta lines show the resulting GWB spectra for 200 random draws from the
posterior parameter distributions of $\talpha$, $\tomega$, and $\tepsilon$. Note
that when constraining the BBH population model, only the leftmost five
frequency-bin data are used, and the $-2/3$ power-law spectrum model has been
assumed. See details in Section~\ref{sec:BBHmodel}.}
\label{fig:GWBConstraint}
\end{figure*}

Figure~\ref{fig:GWBConstraint} shows the NANOGrav 15-year free-spectrum data
(violin symbols) \citep{NG23hd, NG23constraint} and the GWB spectrum (cyan line)
expected from the BBH population model calibrated by the best fit to the
observational data. We denote this BBH model with the best fit of
$(\tilde{\alpha}, \tomega, \tilde{\epsilon})=(1.00, 2.07, 0.30)$ as the
reference BBH population model in this paper. The reference model gives a
magnitude of $A_{\rm yr}=2.0\times 10^{-15}$ at frequency $f=1{\rm yr}^{-1}$,
which is consistent with the value given in \citet[][]{NG23constraint} by
fitting the GWB spectrum with the $-2/3$ power law.

Note that similarly as mentioned above, the requirement of a positively
evolving MBH-host galaxy relation with redshift ($\tomega>0$) is not affected
much by the uncertainty in the adopted MRPG. 
The one-sided uncertainty in the MRPG could be up to a factor of $2-3$
(see the lower panel of Fig.~4 in CYL20).
To consider this, we have done the test by changing the MRPG in the BBH model
to be $2$ or $3$ times larger than that used in the reference model and
matching the expected GWB spectrum to the observed one spectrum, and find
$(\talpha, \tomega, \tepsilon) = (1.00\pm 0.12, 1.45\pm 0.47, 0.31\pm 0.17)$ or
$(1.00\pm 0.12, 1.06\pm 0.49, 0.32\pm 0.17)$. These calculations suggest that
our result on the requirement of a positive evolving MBH-host galaxy relation
is robust. 

\section{Detectability of individual BBHs}
\label{sec:detectability}

With the reference BBH population model, calibrated by the latest NANOGrav
observations, we randomly generate realizations of the cosmic BBHs to calculate
the synthetic strain spectra. We assume that all BBHs are on circular orbits in
the PTA bands \citep{Phinney01}. For a circular BBH system with masses
$M_{\rm BH,1}$ and $M_{\rm BH,2}$ (and total mass
$M_{\rm BBH}=M_{\rm BH,1}+M_{\rm BH,2}$) at redshift $z$, the BBH emits
GWs at a source-rest frequency twice the orbital frequency, i.e., $f\rmr =
2f\orb$, which is then redshifted to $f=f\rmr/(1+z)$. The sky- and
polarization-averaged strain amplitude of the GW from the BBH is 
\begin{equation}
h_0 = \sqrt{\frac{32}{5}}\frac{1}{d_{\rm L}}
\left(\frac{G\calM\rmcz}{c^2}\right)^{5/3}
\left(\frac{\pi f}{c}\right)^{2/3},
\label{eq:h0}
\end{equation}
where $d_{\rm L}$ is the luminosity distance of the BBH system, $\calM\rmcz =
(1+z)\calM\rmcr =
M_{\rm BH,1}^{3/5}M_{\rm BH,2}^{3/5}/M_{\rm BBH}^{1/5}$ are the
redshifted chirp mass. For the GWB produced by BBHs, the synthetic
characteristic strain amplitude at each frequency bin $f_k=k/T\obs$
($k=1,2,...$) is
\begin{equation}
h\rmc(f_k) = \sqrt{\sum_i h_{0,i}^2(f_i)\min(\calN_i,f_iT\obs)},
\label{eq:hc}
\end{equation}
where $\calN=f\rmr^2/\dot{f\rmr}$ and
$\dot{f\rmr} = 96\pi^{8/3}G^{5/3}\calM\rmcr^{5/3}f\rmr^{11/3}/5c^5$.

\begin{figure} 
\centering
\includegraphics[width=0.45\textwidth]{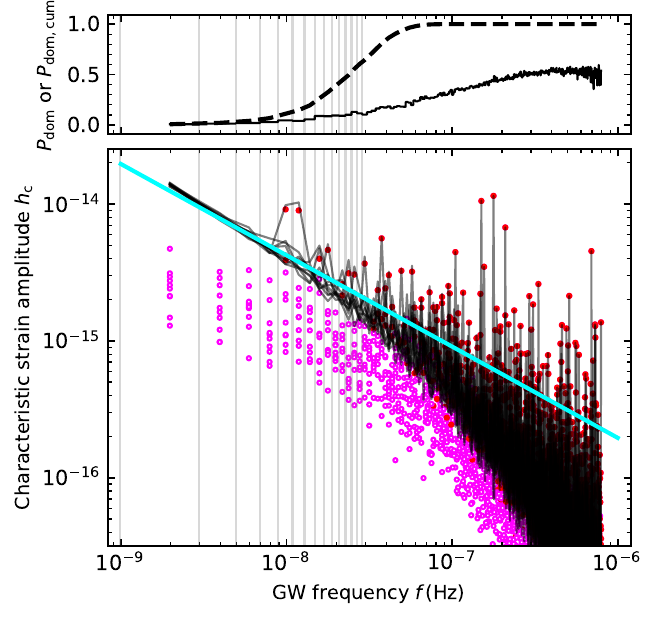}
\caption{Synthetic GWB strain spectra from cosmic BBHs generated from the
reference BBH population model. The lower panel shows the GWB characteristic
strain amplitude $h\rmc$ as a function of the GW frequency $f$ for 10
realizations of the cosmic BBHs (each one represented by a black curve with
significant fluctuation at high frequencies). For comparison, the canonical
$-2/3$ power-law strain spectrum for the same model is indicated by the cyan
line. For each realization, the top contributor to $h\rmc$ within each NANOGrav
frequency bin is marked by a red filled circle if that contributor dominates the
frequency bin, while an open magenta circle if not. In the upper panel, the thin
histogram shows the probability $P_{\rm dom}(f)$ that a given realization
contains a dominant BBH source within a given NANOGrav frequency bin, and the
thick dashed curve shows the cumulative probability $P_{\rm dom,cum}(f)$ that a
given realization contains at least one dominant BBH source in all NANOGrav
frequency bins left to a given frequency. Both probability distributions are
evaluated based on $1000$ realizations generated from the reference BBH
population model. The grey vertical lines in both panels denote the leftmost
$14$ NANOGrav frequency bins. See details in Section~\ref{sec:detectability}.}
\label{fig:GWBSynthesisHigh}
\end{figure}

Figure~\ref{fig:GWBSynthesisHigh} shows the synthetic GWB strain spectra
obtained for $10$ realizations of the BBH population randomly generated from the
reference BBH model (lower panel, each realization represented by a black
curve). For comparison, the canonical $-2/3$ power-law spectrum for the same BBH
model is also shown by the cyan line. As expected, the synthetic spectra are all
well consistent with the canonical power-law spectrum in the left several
NANOGrav frequency bins (marked by the vertical grey lines). At higher
frequencies, however, the synthetic spectra gradually deviate from the canonical
power-law spectrum, with steeper slopes and large fluctuations, due to the
discrete distribution of BBH sources with strong GW signal in different
frequency bins (see \citealt{Sesana08, Roebber16}, CYL20). In some bins, the
spectrum obtained from one realization may be dominated by a single (or a few)
loud individual BBH system(s), which may be detected as individual BBH sources.

For each realization, we record the top contributor to $h\rmc$ within each
NANOGrav frequency bin, and determine if its contribution dominates over the
combined one of the remaining BBH sources within the same frequency bin. We
define the probability that a single BBH dominates the GW radiation in a given
frequency bin $f_k$ as $P_{\rm dom}(f_k)$, and the cumulative probability as
$P_{\rm dom,cum}(f_k)$ that there is at least one BBH source in any of the
frequency bins lower than $f_k$ dominating the GW radiation in that bin. We show
the top contributor in each frequency bin in the lower panel of
Figure~\ref{fig:GWBSynthesisHigh}, marked by the filled red or open magenta
circles if they are the dominant ones or not. In the upper panel, we plot
$P_{\rm dom}(f)$ (thin histogram) and $P_{\rm dom,cum}(f)$ (thick dashed line),
respectively. Both the probability functions are evaluated based on $1000$
realizations of the reference BBH population model. As seen from the figure,
$P_{\rm dom}$ increases nearly monotonically with increasing $f_k$, i.e., the
probability for finding dominant BBH sources is larger at higher frequency bin.
$P_{\rm dom,cum}$ reaches $63\%$ and $95\%$ at the 15th and 27th frequency bins,
corresponding to the frequencies of $30.0\nhz$ and $53.4\nhz$, respectively.
Since the occurrences of dominant sources follow the Poisson distribution, the
occurrence probabilities of $63\%$ and $95\%$ among the realizations correspond
to the expected mean occurrence number of $1$ and $3$ in one realization,
respectively. Therefore, we expect that the probability of finding the signature
induced by dominant BBH sources in the leftmost 15 frequency bins of the
NANOGrav 15-year measurements is larger than $60\%$, if the detected GWB signal
is fully contributed by the cosmic BBHs. We can expect $\sim 1$ dominant source
at $f\lesssim 30\nhz$, and $\sim 3$ at $f\lesssim 54\nhz$.

To investigate the detection of individual BBHs, we consider two sets of PTA
configurations, 
one is NANOGrav with a sensitivity curve represented by the $95\%$ upper limit
on the strain of individual BBHs derived from the NANOGrav 15-year data set
\citep{NG23indv}, and the other is for CPTA with the total number of monitoring
pulsars $N=50$, the timing precision $\sigma_t=100$\,ns, the monitoring cadence
$\Delta t=0.04$\,yr (i.e., about 1 time per 2 weeks).
For CPTA, we first consider the case with an observation
period of $T_{\rm obs}=3.4$\,yr (same as the current CPTA data set). With this
CPTA configuration, we estimate the signal-to-noise ratio ($\snr$) of the
reported HD-correlated GWB according to Equation~(23.69) in
\citet[][]{Maggiore18}, and find that $\snr=4.5$. 
The above CPTA settings (on $N$, $\sigma_t$, $\Delta t$, and $T_{\rm obs}$)
are roughly
consistent with the current CPTA observations (see Fig. 1 in \citealt{CPTA23hd});
and we take them as a surrogate for the current
real one (see more discussion on possible effects of adding noise to this
simple
surrogation below at the end of this section), and estimate its sensitivity
curve for individual BBH detection.
For the details of evaluating the
sensitivity curve of a given PTA on individual sources, we refer to
Section~3.3.2 of \citet{CYL23cgws} (see also
\citealt{Moore15pta, GLY22nf}). 
According to the sensitivity curves, we estimate
the $\snr$ for each mock BBH in each realization. We define those individual
BBHs with $\snr$ larger than a threshold of $\rhoth=3$ or $5$ as detectable ones
that can be resolved from the GWB. Then we calculate the detection probability,
i.e., the fraction of realizations that contains at least one detectable
individual BBHs.

\begin{figure}[!htb]
\centering
\includegraphics[width=0.45\textwidth]{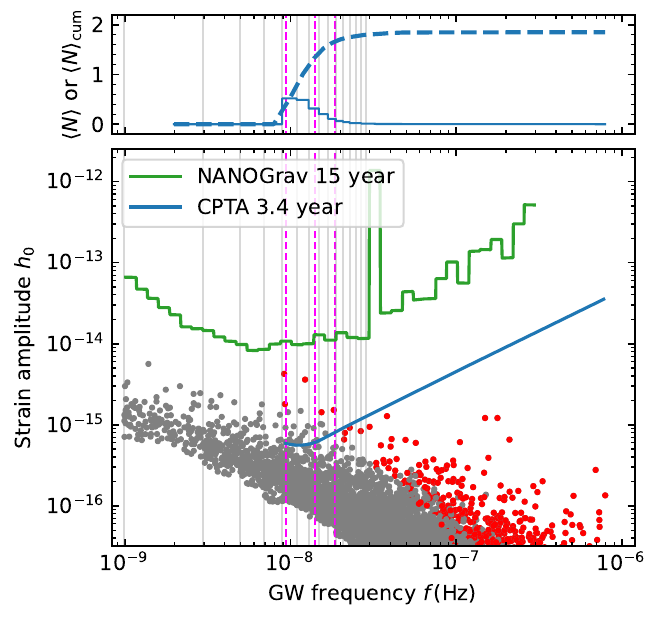}
\caption{Individual BBH sources in the $h_0$--$f$ plane. The lower panel shows
the BBH sources in $10$ realizations generated from the reference BBH population
model. For each realization, we record the top $30$ contributors to the
synthetic GWB characteristic strain amplitude $h\rmc$ within each NANOGrav
frequency bin (filled circles). Among them, those dominating the $h\rmc$ of
their frequency bins are marked in red colors (they are the same sources as
those red filled circles in Fig.~\ref{fig:GWBSynthesisHigh}). The green curve
represents the 95\% upper limit on individual BBHs derived from the NANOGrav
$15$-year data set \citep{NG23indv}, and the blue curve represents the
sensitivity curve on individual sources for the adopted CPTA configuration
assuming an S/N threshold of $3$. In the upper panel, the thin histogram shows
the expected number of individual BBH sources detectable by CPTA in each
NANOGrav frequency bin $\langle N\rangle(f)$, and the thick dashed curve shows
the corresponding cumulative expected number of detectable sources $\langle
N\rangle_{\rm cum}(f)$. Both quantities are the average evaluated based on
$1000$ realizations of cosmic BBHs from the reference BBH population model. The
grey vertical lines in each panel indicate the leftmost $14$ NANOGrav frequency
bins, and the magenta vertical dashed lines denote the three frequencies
explored by CPTA in \citet{CPTA23hd}. See details in
Section~\ref{sec:detectability}.}
\label{fig:LoudConstrained}
\end{figure}

Figure~\ref{fig:LoudConstrained} shows the strain amplitude $h_0$
(see Eq.~\ref{eq:h0}) of individual BBHs from $10$ realizations, together with
both the NANOGrav 15-year sensitivity curve \citep{NG23indv} and the CPTA
sensitivity curve \citep{CYL23cgws}.
The mock BBHs are generated from the reference BBH population model, in which
the detected GWB signal is assumed to be fully from cosmic BBHs. For each
realization, we record the top $30$ contributors to the synthetic GWB
characteristic strain amplitude $h\rmc$ (cf.~Eq.~\ref{eq:hc}) in each NANOGrav
frequency bin (filled circles). Among them, those dominating their frequency
bins are marked by red filled circles (i.e., the same sources as those in
Fig.~\ref{fig:GWBSynthesisHigh}). As seen from the figure, none of the top
contributors is above the 15-year NANOGrav sensitivity curve, suggesting that
current NANOGrav can hardly detect any individual BBH sources; while some
loudest BBHs are already above the sensitivity curve of CPTA which may be
detectable. 
We find that the detection probability of individual BBHs by NANOGrav with
the $15$-year sensitivity curve is $1.5\%$, which is negligible. This result is
consistent with the negative result of searching individual BBHs in the NANOGrav
15-year data set \citep[][similarly for EPTA, \citealt{EPTA23indv}]{NG23indv}.
The detection probability of individual BBHs by CPTA 3.4 yr observations can
be as large as $84.7\%$ (or $64.0\%$) if setting the detection
threshold for detectable BBHs as $\rhoth=3$ (or $5$), with $\sim 1.85$ ($\sim
1.01$) detections of individual BBHs being expected in each realization.
This suggests that there might be resolvable individual BBHs in the current 3.4-year
CPTA data. 

The upper panel of Figure~\ref{fig:LoudConstrained} shows the frequency
distribution of those detectable individual BBH sources expected by CPTA with
the $3.4$\,yr observations. We obtain the expected number of detectable sources
within each NANOGrav frequency bin $\langle N\rangle(f)$ (thin histogram) and
the corresponding cumulative expected number of detectable sources $\langle
N\rangle_{\rm cum}(f)$ (thick dashed curve) according to $1000$ realizations
generated from the reference BBH population model. As seen from this figure,
those detectable sources are most likely to occur at $1$--$2$ frequency bin of
CPTA (i.e., $1$--$2f_0$ where $f_0=\frac{1}{3.4\yr}$). Within the leftmost 14
NANOGrav frequency bins, we find $929$ dominant BBHs in $1000$ realizations from
the reference BBH population model. Among them, $469$ are expected to be
detectable by CPTA with $\snr\geq 3$ within an observation period of $3.4$\,yr.
Thus, the probability of detecting dominant individual BBHs in these frequency
bins by current CPTA is already $\sim 50\%$, for which a careful search of
individual BBHs in the current CPTA data set is strongly motivated. It is worth
to note that most of those detectable BBHs tend to have large total masses
(e.g., $9\lesssim \log_{10}(M_{\rm BBH}/M_\odot)\lesssim 10$), large mass
ratios (e.g., $-1\lesssim \log_{10}\,q_{\rm BBH}\lesssim 0$) and low
redshifts (e.g., $0\lesssim z\lesssim 2$). Most of their signals occur within
frequency range of $\sim 1$--$3\times 10^{-8}\hz$.

For comparison, we also calculate the detection probabilities by setting
different values of the observation period $T\obs$ or the $\snr$ threshold
$\rhoth$, and list them in Table~\ref{tab:t1}. As seen from the table, if
extending the observation time to $T\obs=5\yr$, the detection probability is as
high as $99.5\%$ even for $\rhoth=5$, with $\sim 5.6$ detections being expected.
If extending the observation time to $T\obs=10\yr$, a positive detection can be
guaranteed, i.e., with a detection probability of $99.9\%$ and expected number
of detections $\sim 31.6$ even when we set $\rhoth=5$. 
 
\begin{figure}[!htb]
\centering
\includegraphics[width=0.45\textwidth]{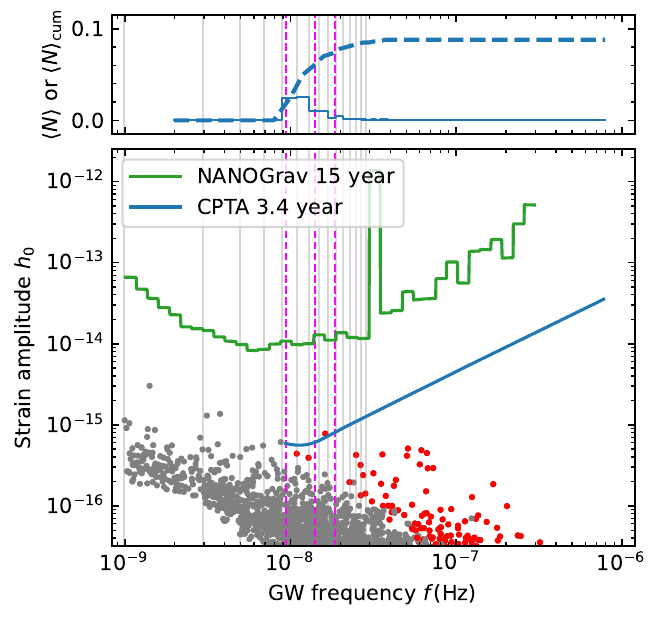}
\caption{Legends are the same as that for Fig.~\ref{fig:LoudConstrained}, except
that the mock BBHs are generated from the BBH population model adopting the
MBH-host galaxy relationship given by \citet{KH13}, which leads to a GWB with
characteristic strain amplitude only a fraction $\sim 28\%$ of the detected GWB
signal. See details in Section~\ref{sec:detectability}.}
\label{fig:LoudEmpirical}
\end{figure}

In the reference BBH population model, we adopt the redshift dependent MBH-host
galaxy scaling relation constrained by the GWB signal by assuming the signal is
fully from the BBH population. It is possible that only a fraction of the signal
is from cosmic BBHs. For example, if the scaling relation is independent of
redshift and the same as the local one given by \citet{KH13}, i.e., with
$(\talpha,\tgamma,\tomega,\tepsilon)=(1.17,8.69,0,0.29)$, the BBH population
model (denoted as the empirical model) would lead to a stochastic background
being only $\sim 28\%$ of the detected GWB signal. In this case, the expected
detection probability and the number of BBH detections should be correspondingly
different from the above estimates for the reference BBH population model. For
comparison, we also generate mock BBHs from the empirical model and estimate the
detection probability and number of BBH detections, as shown in
Figure~\ref{fig:LoudEmpirical} and the right two columns in Table~\ref{tab:t1}.
As seen from the figure and table, CPTA with an observation period of $3.4\yr$
can hardly detect any individual BBHs in this model and the detection
probability is only about $8.5\%$ assuming $\rhoth=3$. If extending the
observation period to $T\obs=5\yr$ or $10\yr$, then the expected number of BBH
detections is about $0.539$ or $10.4$ with $\rhoth=3$, and correspondingly the
detection probability is about $41.5$\% or $100\%$. These detection numbers are
substantially less than those expected from the reference BBH population model,
which suggests that the detection of individual BBHs by PTA experiments can be
used jointly with the GWB signal to put constraints on the BBH population model
and the fraction of the GWB signal contributed by cosmic BBHs. 

\begin{deluxetable*}{cCCCCC}
\tablecaption{Detection probabilities and corresponding expected numbers of
detectable individual BBHs (in the bracket) from the BBH population model by CPTA with
different settings of $T\obs$ and $\rhoth$. The BBH population model adopts
either the MBH--host galaxy scaling relation calibrated by the GWB signal
reported by NANOGrav (reference Model) or the empirical one given by
\citet{KH13} (empirical Model). \label{tab:t1}}
\tablehead{\multirow{2}{*}{$T\obs(\yr)$} & \multicolumn{2}{c}{Reference Model}
&& \multicolumn{2}{c}{Empirical Model} \\ \cline{2-3} \cline{5-6}
& \dcolhead{\rhoth=3} & \dcolhead{\rhoth=5} && \dcolhead{\rhoth=3} &
\dcolhead{\rhoth=5}}
\startdata
3.4 & 84.7\%\,(1.85) & 64.0\%\,(1.01) && \phn8.5\%\,(0.088) & \phn5.0\%\,(0.052)\\
5.0 & \phd100\%\,(10.1) & 99.5\%\,(5.59) && 41.5\%\,(0.539) & 26.8\%\,(0.311)\\
10 & \phd100\%\,(34.2) & \phd100\%\,(28.3) && \phd100\%\,(10.4)\phn & 99.9\%\,(6.34)\phn\\
\enddata
\end{deluxetable*}

We then investigate in detail how a non-detection of individual BBHs by CPTA in
the near future, if it was, can be converted into the constraint on the
contribution of cosmic BBHs to the GWB. Assuming $\calA_{\rm yr}$ and
$\calA_{\rm yr,BBH}$ represent the characteristic strain amplitude of the
detected GWB and that of the GWB induced by the cosmic BBHs at the reference
frequency $f=1\iyr$, respectively, we define the ratio $\calH_{\rm BBH}\equiv
\calA_{\rm yr,BBH}/\calA_{\rm yr}$ to indicate the significance of the
contribution from the cosmic BBHs to the total GWB signal. With this definition,
the BBH-induced and non-BBH-induced components contribute a fraction of
$\calH_{\rm BBH}^2$ and $1-\calH_{\rm BBH}^2$ to the total power of the detected
GWB, respectively. By assuming any given $\calH_{\rm BBH}$, we can calibrate the
BBH population synthesis model by adjusting the mass scaling relation
(Eq.~\ref{eq:scaling}) as described in Section~\ref{sec:BBHmodel} and then
generate mock BBHs according to the calibrated model. With the mock sample, we
can calculate the detection probability of individual BBHs by CPTA with any
given observational period $T\obs$. Then we can estimate the constraint on
$\calH_{\rm BBH}$ if none of the cosmic BBHs was detected by CPTA with an
observational period of $T\obs$ as described in the previous paragraph.

\begin{figure}[!htb]
\centering
\includegraphics[width=0.45\textwidth]{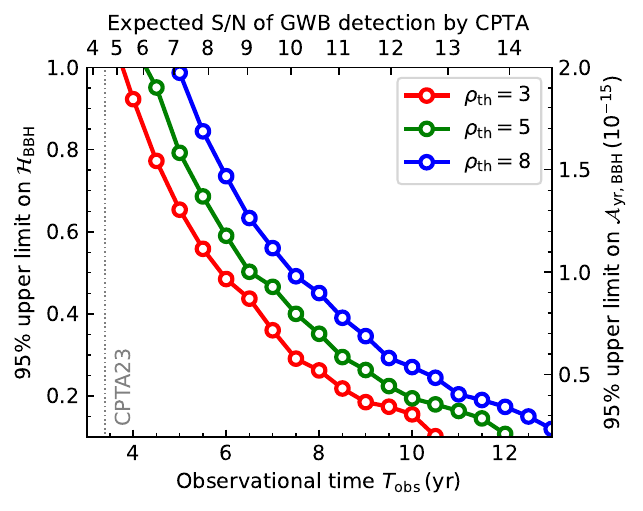}
\caption{Expected constraints ($95\%$ upper limits) on the ratio of the 
BBH-induced GWB strain amplitude to the detected GWB strain amplitude
$\calH_{\rm BBH}$, as a function of the CPTA observation time $T\obs$, if there
is no detection of individual BBHs within $T\obs$. Red, green, and blue colors
correspond to the cases in which the threshold $\snr$ for individual BBH
detections are $\rhoth=3$, $5$, and $8$, respectively. The corresponding $\snr$
of the GWB detection by CPTA is shown by the top axis, and the BBH-induced
characteristic strain amplitude at frequency $f=1\iyr$ is shown by the right
axis. The vertical dotted line marks the $3.4\yr$ CPTA observation. See details
in Section~\ref{sec:detectability}.}
\label{fig:LoudFraction}
\end{figure}

Figure~\ref{fig:LoudFraction} shows the inferred $95\%$ upper limit on
$\calH_{\rm BBH}$ as a function of the CPTA observation time $T\obs$ (or the GWB
detection $\snr$) by assuming that none of individual BBHs could be detected by
CPTA within $T\obs$. That is, at a given $T\obs$, the detection probability of
individual BBHs by CPTA should be greater than $95\%$ if $\calH_{\rm BBH}$ is
above the value indicated by the curves in the figure. As seen from the figure,
the contribution of the cosmic BBHs to the detected GWB can be effectively
constrained at $T\obs\gtrsim 4$--$5\yr$, or equivalently, when the GWB detection
has $\snr\gtrsim 5$--$7$, jointly by the GWB signal and the
detection/non-detection of individual BBHs. If non-detection is made by CPTA
with $T\obs\sim 6$--$7.5\yr$ (with GWB detection $\snr\sim 8$--$10$),
$\calH_{\rm BBH}$ should be below $0.5$, and the contribution fraction of cosmic
BBHs to the total power of the detected GWB signal should be $\lesssim 25\%$. If
the CPTA does not detect individual BBHs with $T\obs\sim 7.5$--$9.5\yr$ (or with
GWB detection $\snr\sim 10$--$12$), $\calH_{\rm BBH}$ should be below $0.3$,
which thus suggests the GWB strain amplitude produced by cosmic BBHs should be
below that predicted by the empirical model.

Adopting a similar BBH population
synthesis model, in which the MBH--host galaxy scaling relation is the same as
the local one given by \citet{KH13} without redshift evolution, as an example,
the GWB induced by the cosmic BBHs may be a factor $\sim 0.3$ of the
detected GWB strain signal. In this case, our calculations show that a $3.4$-year accumulation of the CPTA
data leads to a detectability of individual BBHs $\lesssim 8.5\%$, and that a
$7.4$-year (or $8.1$-year) accumulation of the data is needed for CPTA to have a
detection probability of $95\%$ (or $99\%$).

Note that the above detection probabilities and numbers for CPTA are obtained
based on an idealized estimate for the sensitivity curve (the blue curve
in Figs.~\ref{fig:LoudConstrained} and \ref{fig:LoudEmpirical}) without
considering some complicate factors (e.g., unmodeled red noises or more
complicated noises) that may affect the sensitivity. 
These factors may induce special features to the sensitivity curve at some
specific frequencies, and therefore affect the detection of individual GW
sources at these frequencies. Given the fact that such effects are hard to
quantify for a mock PTA configuration, we simply assume a conservative case, in
which the actual sensitivity for individual source detection is about a factor
of $3$ times worse than the idealized estimation presented above.
In that assumed case, for a detection threshold of $\rhoth=3$, we find that the
detection probability resulting from the reference model is $11.9\%$ for
$T\obs=3.4\yr$.
A detection probability of $95\%$ (or $99\%$) is expected for the reference
model if extending the observation period to $T\obs=6.7\yr$ (or $7.3\yr$);
while such a detection probability from the empirical model requires an
observation time period of $12.6\yr$ (or $13.8\yr$). 
Even in the above assumed conservative case, it is still plausible to conclude
that the breakthrough of individual BBH detection should be realized within a
few years or about ten years, and strong constraints on the BBH and/or
cosmological origin of the GWB can be obtained in the near future. 

\section{Conclusions}
\label{sec:Conclusions}

In this work, we investigate the implications of the recently reported evidence
for a stochastic GWB by several PTAs \citep{NG23hd, EPTA23hd,PPTA23hd, CPTA23hd}
on the detection of individual BBHs. We first adopt the reported GWB signal to
constrain the BBH population synthesis model, focusing specifically on the
redshift evolution of the MBH--host galaxy scaling relation, by assuming that
the GWB signal is entirely contributed by cosmic BBHs. With the constrained
model, we then generate random realizations of the cosmic BBH populations to
study the detections of individual BBHs as well as the dominance of individual
BBHs over the stochastic GWB in different GW frequency bins. The detected GWB
implies a significant positive redshift evolution of the MBH--host galaxy
scaling relation, which is consistent with \citet{CYL23cgws}. We find a
considerable probability that there have already been signatures emerged in the
current NANOGrav 15-year free-spectrum data set as caused by some individual
BBHs dominating their frequency bins. Their occurrence probabilities are $63\%$
and $95\%$ within the leftmost $15$ and $27$ NANOGrav frequency bins,
corresponding to $f\leq 30\nhz$ and $f\leq 54\nhz$, respectively. However, given
the current NANOGrav's capability \citep{NG23indv}, the probability of detecting
these sources individually is rather low (i.e., $\lesssim 2\%$). We further find
that those loudest BBHs, if any, in the current CPTA data set ($3.4$\,yr) may be
detectable with a detection probability of $\sim 85\%$ for a detection S/N
threshold of $3$, and with $1.85$ such detectable BBHs being expected. If
extending the observation time of CPTA to $5\yr$, a positive detection of
individual BBHs is almost guaranteed, i.e., successful detection in each
realization, with the mean detection number of $\sim 10$ expected in each. 

If the cosmic BBHs only contribute a fraction of the detected signal, the
evolution of the MBH-host galaxy scaling relation may not be required and the
detection of individual BBHs is less likely.  However, if the contribution from
cosmic BBHs to the total power of the detected GWB signal is $\gtrsim 10\%$ (or
$\calH_{\rm BBH}\gtrsim 0.3$), a positive detection of individual BBHs by CPTA
can still be expected with an observation period of $\sim 10\yr$.

Jointly with the detected GWB signal, even if no individual BBHs are detected by
CPTA in the coming several years, the non-detection can be converted to the
constraint on the upper limit of the cosmic BBH contribution to the detected
GWB. For example, the non-detection of individual BBHs by CPTA with $T\obs \sim
6$--$7.5\yr$ or $7.5-9.5\yr$ suggests that the ratio of the BBH-induced GWB
strain amplitude to the detected GWB strain amplitude ($\calH_{\rm BBH}$) is
below $0.5$ or $0.3$, and the contribution fraction of the BBHs to the total
power of the detected GWB signal is below $25\%$ or $10\%$. We conclude that the
detection (or non-detection) of individual BBHs by CPTA in the coming several
years can play an important role in not only interpreting the astrophysical
origin of the recently detected GWB signal but also putting strong constraint on
the contribution from the cosmic BBHs to the GWB signal.

\section*{acknowledgements}
This work is partly supported by the National SKA Program of China (grant no.
2020SKA0120101), National Key Program for Science and Technology Research and
Development (grant nos. 2022YFC2205201, 2020YFC2201400), and the National
Natural Science Foundation of China (grant nos. 12173001, 12273050, 11721303,
11991052). 

\end{document}